\documentclass{article}

\usepackage{arxiv}

\usepackage[utf8]{inputenc} 
\usepackage[T1]{fontenc}    
\usepackage{hyperref}       
\usepackage{url}            
\usepackage{booktabs}       
\usepackage{amsfonts}       
\usepackage{nicefrac}       
\usepackage{microtype}      
\usepackage{lipsum}		
\usepackage{graphicx}
\usepackage{natbib}
\usepackage{doi}
\usepackage{amsmath}
\usepackage{tabularx}
\usepackage{comment}

\title{MiTU-Net: A fine-tuned U-Net with SegFormer backbone for segmenting pubic symphysis-fetal head}

\date{} 					

\author{ \href{https://orcid.org/0009-0003-0427-368X}{\includegraphics[scale=0.04]                    {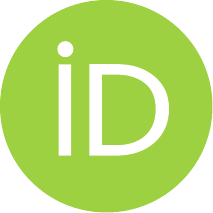}\hspace{0.5mm}Fangyijie Wang} \\
	School of Medicine\\
    University College Dublin\\
	\texttt{fangyijie.wang@ucdconnect.ie} \\
	\And
        \hspace{0.5mm}Gu\'enol\'e Silvestre \\
	School of Computer Science\\
    University College Dublin\\
	\texttt{guenole.silvestre@ucd.ie} \\
        \And
        \href{https://orcid.org/0000-0003-0095-9337}{\includegraphics[scale=0.04]{orcid.pdf}\hspace{0.5mm}Kathleen Curran} \\
	School of Medicine\\
    University College Dublin\\
	\texttt{kathleen.curran@ucd.ie} \\        
}

\renewcommand{\shorttitle}{Methodology}

\hypersetup{
pdftitle={MiTUNet: A fine-tuned U-Net with SegFormer backbone for segmenting pubic symphysis-fetal head},
pdfsubject={q-bio.NC, q-bio.QM},
pdfauthor={David S.~Hippocampus, Elias D.~Striatum},
pdfkeywords={First keyword, Second keyword, More},
}

\begin{document}
\maketitle

\begin{abstract}

Ultrasound measurements have been examined as potential tools for predicting the likelihood of successful vaginal delivery. The angle of progression (AoP) is a measurable parameter that can be obtained during the initial stage of labor. The AoP is defined as the angle between a straight line along the longitudinal axis of the pubic symphysis (PS) and a line from the inferior edge of the PS to the leading edge of the fetal head (FH). However, the process of measuring AoP on ultrasound images is time consuming and prone to errors. To address this challenge, we propose the Mix Transformer U-Net (MiTU-Net) network, for automatic fetal head-pubic symphysis segmentation and AoP measurement. The MiTU-Net model is based on an encoder-decoder framework, utilizing a pre-trained efficient transformer to enhance feature representation. Within the efficient transformer encoder, the model significantly reduces the trainable parameters of the encoder-decoder model. The effectiveness of the proposed method is demonstrated through experiments conducted on a recent transperineal ultrasound dataset. Our model achieves competitive performance, ranking 5th compared to existing approaches. The MiTU-Net presents an efficient method for automatic segmentation and AoP measurement, reducing errors and assisting sonographers in clinical practice. Reproducibility: Framework implementation and models available on https://github.com/13204942/MiTU-Net.

\end{abstract}


\section{Introduction}

\begin{figure}
    \centering
    \includegraphics[width=0.8\linewidth]{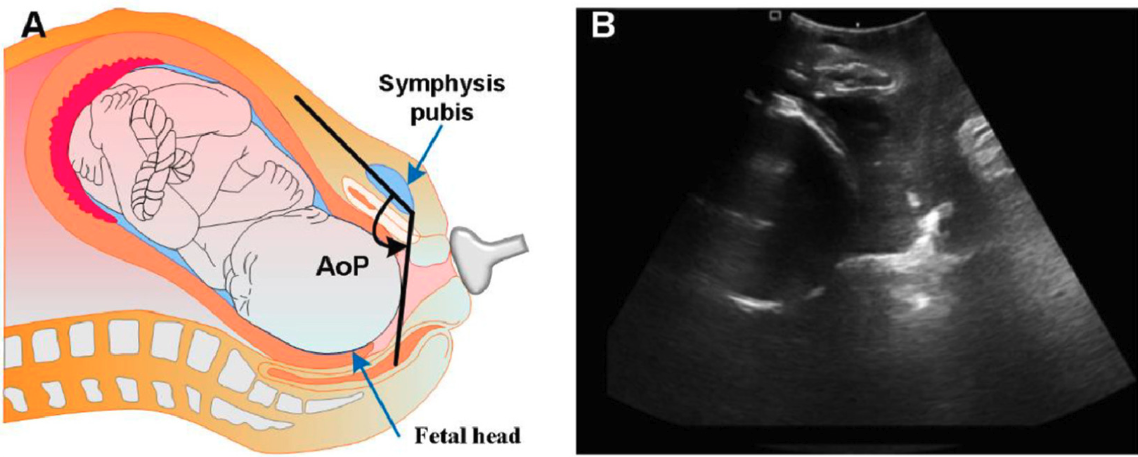}
    \caption{The illustration of the process of measuring the angle of progression (AoP) using transperineal ultrasound. (A) Schematic diagram of calculating AoP. (B) An image showing the symphysis pubis and fetal head.}
    \label{fig:fh_ps}
\end{figure}

Assessing the progression of labor is vital to ensure proper advancement in peripartum care. This assessment enables early detection of any deviations from the expected course, aiming to minimize potential complications for both the mother and fetus. The International Society of Ultrasound in Obstetrics and Gynecology (ISUOG) recommends the implementation of ultrasound biometric technology, particularly the angle of progression (AoP), when assessing the fetal craniopubic symphysis structure~\cite{Ghi:2018}, shown in Figure~\ref{fig:fh_ps}. Thus, the measurement of AoP proves to be a superior method for assessing labor progression. AOP is defined as the angle between a straight line extending along the longitudinal axis of the pubic symphysis (PS) and a line connecting the inferior edge of the PS to the leading edge of the fetal head (FH)~\cite{Lu:2022}. The current measurement of AoP is manual assessments conducted by experienced physicians. However, this process is time-consuming, cumbersome, and prone to measurement errors. To address this problem, an algorithm for the automatic measurement of AoP is highly demand in clinical practice. This contribution improves the accuracy of labor progress assessment. The accurate segmentation of the FH and PS regions from ultrasound images is an essential prerequisite for the automated measurement of AoP.

In recent years, deep learning (DL) technology experienced significant advancements. In the field of medical imaging analysis, convolutional neural networks (CNNs), a type of deep learning technology, have demonstrated impressive performance in a wide range of medical imaging applications. Three notable architectures are fully convolutional networks (FCNs)~\cite{Long:2015}, U-Net~\cite{Ronneberger2015}, and the three-dimensional V-Net~\cite{Milletari:2016}. These architectures have gained recognition as classic models for analyzing medical images.

Recently, researchers have made attempts to combine CNN and Transformer~\cite{Vaswani:2017} architectures for the purpose of medical image segmentation.~\cite{Chen:2021} propose to combine U-Net and Transformer to obtain TransUnet for medical image segmentation. To analysis pregnancy ultrasound images with automatic DL algorithms,~\cite{Chen:2024} propose a model incorporating a dual attention module, a multi-scale feature screening module and a direction guidance block for segmenting the fetal head and pubic symphysis. This model offers significant advantages in extracting global contextual information and refining segmentation results with direction guidance.
 
This paper introduces an efficient segmentation network based on the encoder-decoder framework, U-Net, for the automated segmentation of ultrasound images and the measurement of AoP. We enhance the U-Net architecture by replacing its encoder path with a pre-trained Mix Transformer~\cite{xie2021}. The Mix Transformer encoder includes attention mechanisms to enhance feature representation. Furthermore, it reduces computational cost and memory usage by selecting and processing only important parts of the input. Additionally, it significantly reduces the number of trainable parameters from 31 million to 5 million in comparison to the traditional U-Net.

The rest of this paper is organized as follows. Section~\ref{sec:method} presents the detailed description of the proposed model. Section~\ref{sec:results} focuses on the analysis and evaluation of the experimental results.

\section{Methodology}
\label{sec:method}

\begin{figure}
    \centering
    \includegraphics[width=1\linewidth]{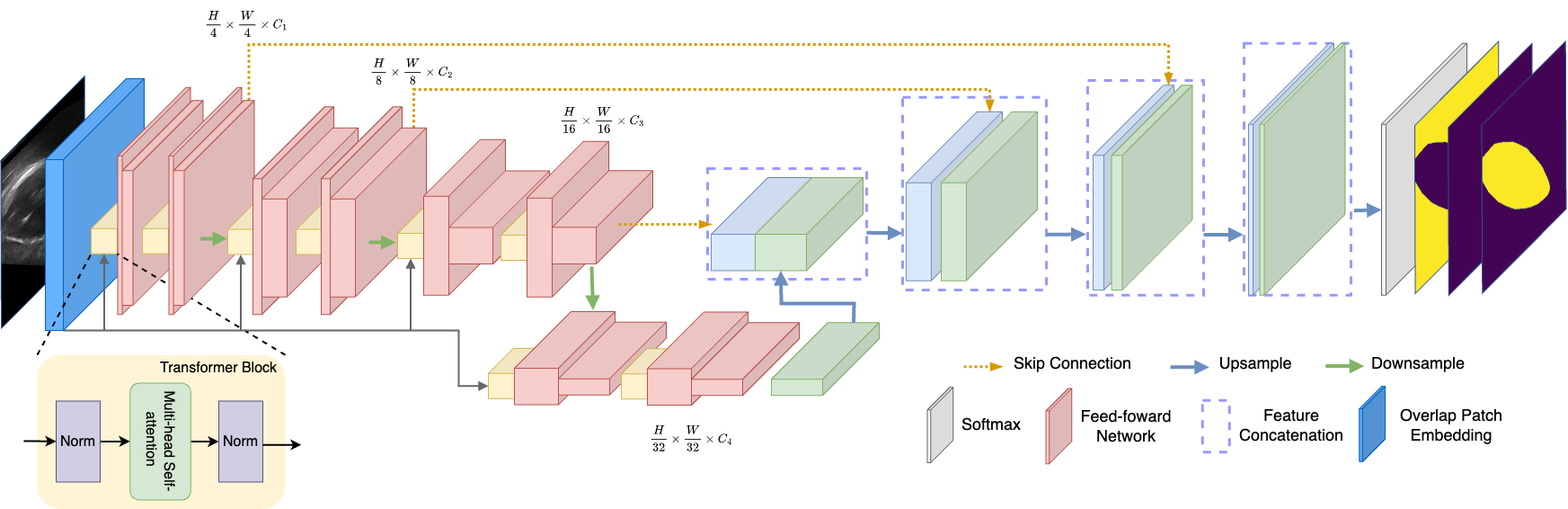}
    \caption{The illustration of the proposed MiTU-Net for automatic fetal head (FH)-pubic symphysis (PS) segmentation.}
    \label{fig:unet_view}
\end{figure}

This section introduces an efficient and powerful segmentation fine-tuned model with low computationally demanding modules, MiTU-Net, see Figure~\ref{fig:unet_view}. The model consists of a U-Net architecture with a pre-trained Mix Transformer encoder for segmenting pubic symphysis-fetal head on a ultrasound dataset, JNU-IFM~\cite{lu2022ifm}. The U-Net architecture is followed the design from Ronneberger~\cite{Ronneberger2015} in 2015. The contracting path ("encoder") of the U-Net is replaced with a pre-trained Mix Transformer encoder (MiT-B0) designed by Xie in SegFormer framework~\cite{xie2021} in 2021. The framework is developed by programming language Python, library PyTorch, and library Segmentation Models~\cite{Iakubovskii2019}.

\subsection{Model Achitecture}
Our U-Net architecture contains the contracting path ("encoder") and the expansive path ("decoder"). They both have the same depth of four layers. Between encoder and decoder, we use skip connections to concatenating features from the Mix Transformer encoder to recover spatial information lost during downsampling. We first transpose the original ultrasound image of size $3 \times 256 \times 256$ to size of $256 \times 256 \times 3$. Then the Mix Transformer encoder divides the ultrasound image into small patches of size $4 \times 4$. These patches are the input to the Mix Transformer encoder to obtain multi-level features at ${\{1/4, 1/8, 1/16, 1/32\}}$ of the original image resolution. The encoder passes the multi-level features to the decoder to predict the segmentation mask at a $\frac{H}{4} \times \frac{W}{4} \times N_{cls}$ resolution, where $N_{cls}$ is the number of categories. The JNU-IFM dataset contains three categories labeled as 0, 1, or 2, with 0 representing the background, 1 representing the pubic symphysis, and 2 representing the fetal head.

\paragraph{Feature Representation}
The Mix Transformer encoder is a hierarchical type of Transformer encoder that produces multi-level features. It provides both high-resolution coarse features and low-resolution fine features. Specifically, the encoder creates a hierarchical feature map consisting of a resolution of $\frac{H}{2^{i+1}} \times \frac{W}{2^{i+1}} \times C_i$, where $i \in\{1,2,3,4\}$, by merging multiple patches extracted from the input ultrasound image.

\paragraph{Overlapped Patch Merging}
We reproduce the Overlapped Patch Merging (OPM) process proposed by Xie~\cite{xie2021} in the patch merging process for SegFormer. It requires $K, S,$ and $P$, where $K$ is the patch size, $S$ is the stride between two adjacent patches, and $P$ is the padding size. In the first OPM step, we set $K=7, S=4, P=3$, and $K=3, S=2, P=1$ for the rest of OPM steps to perform patch merging to produces features.

\paragraph{Self-Attention}
We apply the multi-head self-attention process~\cite{Vaswani2017}, each of the heads $Q, K, V$ have the same dimensions $H \times W \times C$, where $H$ is the height of input image,  $W$ is the width of input image, and $C$ is the number of channels. The self-attention is estimated as:
\begin{equation} \label{eq1}
\operatorname{Attention}(Q, K, V)=\operatorname{Softmax}\left(\frac{Q K^{\mathrm{\top}}}{\sqrt{d_{head }}}\right) V 
\end{equation}
The computational complexity of this process is $O\left(N^2\right)$,

\paragraph{Feed-forward Network}
We use a feed-forward network (FFN) after the the self-attention module. This FFN contains a $3 \times 3$ depth-wise convolution that can provide positional information for Transformers sufficiently and reduce computational cost~\cite{xie2021}. It is formulated as follows:
\begin{equation} \label{eq2}
\begin{split}
\mathbf{\hat{x}}_{out} &= \operatorname{Linear}(C_{in}, C_{out})(\mathbf{x}_{in}) \\
\mathbf{x}_{out} &= \operatorname{Linear}(C_{out}, C_{in})(\operatorname{GELU}(\operatorname{DWConv}_{3 \times 3}(\mathbf{\hat{x}}_{out}))) \\
\end{split}
\end{equation}
where $\mathbf{x}_{in}$ is the feature map from the self-attention module, and Linear $\left(C_{\text {in }}, C_{\text {out }}\right)(\cdot)$ is a linear layer with $C_{in}$ and $C_{out}$ as input and output vector dimensions respectively.  $\operatorname{DWConv}_{3 \times 3}$ refers to the $3 \times 3$ depth-wise convolution.

\subsection{Decoder}
The decoder path is symmetric to the encoder path. In total it has four convolutional layers. Every layer consists of an upsampling of the feature map followed by a $2\times2$ up-convolution that halves the number of feature channels, a concatenation with the correspondingly cropped feature map from the encoder path, and two $3\times3$ convolutions, each followed by a ReLU. The bottom of decoder path is a single $3\times3$ ("segmentation head") up-convolution followed by a Softmax that outputs asegmentation map with 3 channels. We formulate the decoder's single layer as:
\begin{equation} \label{eq3}
\begin{split}
\hat{X}_{out} &= {\operatorname{Conv}_{3\times3}}(C_{F}, C_{out})(F) \\
\hat{X}_{out} &= {\operatorname{ReLU}}({\operatorname{BatchNorm2d}}(\hat{X}_{out})) \\
X_{out} &= {\operatorname{Conv}_{3\times3}}(C_{out}, C_{out})(\hat{X}_{out}) \\
X_{out} &= {\operatorname{ReLU}}(\operatorname{BatchNorm2d}(X_{out})) \\
\end{split}
\end{equation}

where $F$ refers to the feature maps from Mix Transformer encoder path concatenated with the upsampling feature maps from decoder path, $C_{F}$ refers to the dimensions of the concatenated feature maps $F$, and $C_{out}$ is the output vector dimensions.

\subsection{Data pre-processing and augmentation}
The JNU-IFM dataset has 4000 ultrasound images in total. All of them are used for training. We randomly select 30 ultrasound ultrasound images for validation during training. The size of images is $3 \times 256 \times 256$. We apply one hot encoding on ground truth masks for semantic segmentation, accordingly the ground truth prediction has 3 channels, background prediction, pubic symphysis prediction, and fetal head prediction.

Initially, we are inspired by the regularization technique of randomly masking out square regions of input during training~\cite{devries2017}, therefore, we randomly mask out $4 \times 4$ square regions of the input images. The number of masked regions is in range $[0, 4]$. Additionally, we pick a random angle from $[-25^{\circ}, 25^{\circ}]$ to rotate the input image. Also, we randomly flip the input image horizontally with 50\% probability, and vertically with 30\% probability. All of input pixels are normalized with mean values [0.0, 0.0, 0.0], and standard deviation values [1.0, 1.0, 1.0]. The validation images have no augmentation, but their input pixels are also normalized with mean values [0.0, 0.0, 0.0], and standard deviation values [1.0, 1.0, 1.0].

\subsection{Fine-tuning}
We implemented all of our experiments using PyTorch, then we trained the entire U-Net with pre-trained MiT-B0 encoder with 20 epochs from scratch. The training dataset and validation dataset both have a batch size of 10. The Adam optimiser was used in training processes with a learning rate of \(1e-4\). All training processes were performed on an NVIDIA Tesla T4 graphics card. 

\paragraph{Loss Function}
During the training process, we construct the loss function for this segmentation task with the Binary Cross Entropy loss function and Jaccard Index. Given a predicted mask $\hat Y$ and a ground truth mask $Y$, the loss function can be formulated as follows:
\begin{equation} \label{eq4}
\mathcal{L}= \operatorname{BCE}(Y, \hat Y) + (1 - \operatorname{J}(Y, \hat Y))
\end{equation}
where $\operatorname{BCE}$ refers to the Binary Cross Entropy loss function, and $\operatorname{J}$ refers to the Jaccard Index (or Intersection over Union).

\paragraph{Evaluation Metrics}
The typical metrics applied to evaluate the performance of this U-Net architecture are Pixel Accuracy (PA), Dice coefficient~\cite{Dice:1945}, and Mean Intersection over Union (IoU)~\cite{Jaccard:1912}. Mean IoU is defined as the average IoU over all classes \(K\). These metrics can be computed using true positive(TP), false positive (FP), true negative (TN) and false negative (FP) of the segmentation results~\cite{Tavakoli:2021}. TP represents the pixels considered as being in the object and being really in the object, conversely, TN is the pixels outside the object both in the segmentation and the ground truth. FP represents the pixels considered by the segmentation in the object, but which in reality are not part of it. Finally, FN represents the pixels of the object that the segmentation has classified outside.

\section{Experiments Results}
\label{sec:results}

\subsection{Evaluation Metrics}

The typical metrics applied to evaluate the performance of segmentation models are Dice similarity coefficient (DSC)~\cite{Dice:1945}, Hausdorff Distance (HD)~\cite{Birsan:2006}, Average Surface Distance (ASD)~\cite{Yeghiazaryan:2018}, and the AoP difference ($\Delta$AoP) between predicted and manually measured AoP for prediction. The DSC is employed to assess the level of similarity between a predicted segmentation mask and the ground truth segmentation mask. The DSC values range from 0, representing no overlap, to 1, indicating perfect overlap. HD measures the maximum spatial dissimilarity between corresponding points in the predicted and ground truth masks. On the other hand, the ASD quantifies the average spatial dissimilarity between the predicted and ground truth masks, offering a different perspective on segmentation accuracy. The final evaluation metrics can be defined by~\ref{eva_metric}:

\begin{equation} 
\label{eva_metric}
\begin{split}
S &= 0.25\left(\frac{D S C_{F H}+D S C_{P S}+D S C_{A L L}}{3.0}\right) \\
& + 0.25\left[\frac{0.5\left(1-\frac{H D_{F H}}{100}+1-\frac{H D_{P S}}{100}+1-\frac{H D_{A L L}}{100}\right)}{3.0}+\frac{0.5\left(1-\frac{A S D_{F H}}{100}+1-\frac{A S D_{P S}}{100}+1-\frac{A S D_{A L L}}{100}\right)}{3.0}\right] \\
& + 0.5\left(1-\frac{\Delta A o P}{180}\right)
\end{split}
\end{equation}

\begin{figure}[htbp]
    \centering
    \includegraphics[width=1\linewidth]{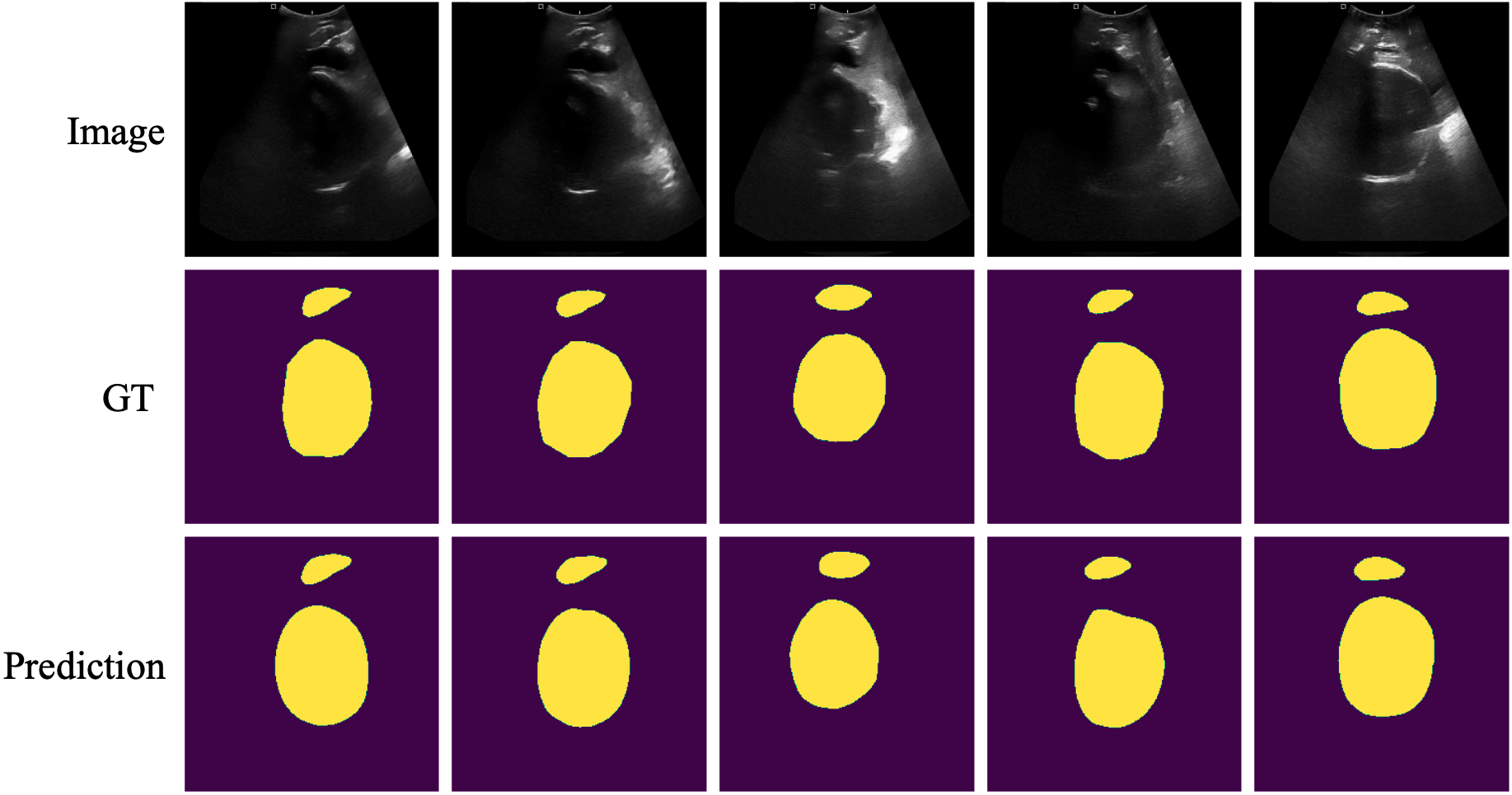}
    \caption{Visualization of segmentation results on the JNU-IFM dataset. GT: Ground Truth}
    \label{fig:vis_seg}
\end{figure}

Besides, the visualization of the segmentation results obtained by MiTU-Net is shown in Figure~\ref{fig:vis_seg}. The second row displays the ground truth masks, while the third row presents the segmentation details obtained from MiTU-Net. The results indicate a competitive performance of MiTU-Net in segmenting the pubic symphysis and fetal head shapes.

\subsection{Compare with SOTA Methods}

Compared to other methods, our proposed method achieved the 5th place in the FH-PS-AOP challenge, shown in Table~\ref{tab:numeric_results}. The MiTU-Net achieves a final score of 0.9283, demonstrating its effectiveness. Moreover, its average Dice score of 0.9247 for the fetal head and public symphysis demonstrates its competitiveness, surpassing even the rank 1 method. However, it is noteworthy that our method has big gap between top-ranking methods on HD for fetal head and pubic symphysis. We believe the reason for this is that HD is not considered as a loss during fine-tuning process. 

\begin{table}[htbp]
    \caption{The final segmentation results of top 7 methods evaluated on the FH-PS-AOP dataset.}
    \renewcommand{\arraystretch}{1.3}
    \begin{tabularx}{\linewidth}{
        >{\arraybackslash}p{0.03\linewidth}
        >{\centering\arraybackslash}p{0.05\linewidth}
        >{\centering\arraybackslash}p{0.05\linewidth}
        >{\centering\arraybackslash}p{0.06\linewidth}
        >{\centering\arraybackslash}p{0.06\linewidth}
        >{\centering\arraybackslash}p{0.06\linewidth}
        >{\centering\arraybackslash}p{0.05\linewidth}
        >{\centering\arraybackslash}p{0.05\linewidth}
        >{\centering\arraybackslash}p{0.05\linewidth}
        >{\centering\arraybackslash}p{0.05\linewidth}
        >{\centering\arraybackslash}p{0.05\linewidth}
        >{\centering\arraybackslash}p{0.05\linewidth}
        >{\centering\arraybackslash}p{0.05\linewidth}
        }
        \hline
        \centering\textbf{Rank} & \textbf{Author} & \textbf{Final Score} & \textbf{AOP} & \textbf{HD FH} & \textbf{HD PS} & \textbf{HD All} & \textbf{ASD FH} & \textbf{ASD PS} & \textbf{ASD All} & \textbf{Dice FH} & \textbf{Dice PS} & \textbf{Dice All} \\
        \specialrule{1.5pt}{1pt}{1pt}
        \centering 1 & Stock & \textbf{0.9418} & 6.5437 & 12.6313 & 7.6385 & 13.4477 & 3.8963 & \textbf{2.4086} & 3.4857 & 0.9303 & 0.8833 & 0.9236 \\
        \centering 2 & Elbatel & 0.9416 & 7.9698 & \textbf{10.6985} & \textbf{7.5586} & \textbf{12.0595} & \textbf{3.3069} & 2.9945 & \textbf{2.9811} & \textbf{0.9403} & \textbf{0.8881} & \textbf{0.9346} \\
        \centering 3 & Qiu & 0.9388 & 7.6472 & 12.4586 & 7.6612 & 13.6147 & 3.6155 & 2.2572 & 3.2383 & 0.9358 & 0.8696 & 0.9295 \\
        \centering 4 & Chen & 0.9312 & 8.5581 & 14.0105 & 9.0512 & 15.3342 & 3.8692 & 2.6199 & 3.5173 & 0.9314 & 0.8596 & 0.9240 \\
        \centering 5 & \textbf{Ours} & 0.9283 & 8.7188 & 14.0093 & 10.8286 & 15.8089 & 3.9837 & 2.9824 & 3.5785 & 0.9313 & 0.8580 & 0.9247 \\
        \centering 6 & Sun & 0.9230 & 9.2760 & 15.7951 & 11.5361 & 17.5598 & 4.7231 & 3.1139 & 4.2646 & 0.9179 & 0.8370 & 0.9096 \\
        \centering 7 & Cai & 0.8974 & 12.1990 & 20.0307 & 14.0684 & 21.8731 & 7.0988 & 4.2081 & 6.0576 & 0.8787 & 0.8036 & 0.8725 \\
        \hline
    \end{tabularx}
    \label{tab:numeric_results}
\end{table}

\section*{Acknowledgments}
This publication has emanated from research conducted with the financial support of Science Foundation Ireland under Grant number 18/CRT/6183. For the purpose of Open Access, the author has applied a CC BY public copyright licence to any Author Accepted Manuscript version arising from this submission.

\bibliographystyle{abbrvnat}
\bibliography{references}

\end{document}